\documentclass[12pt]{iopart}

\usepackage{graphicx}
\usepackage{iopams}
\usepackage{matutil}
\usepackage{subfigure}
\usepackage{harvard}

\begin{document}

\title{Using a phase space cross section to study large complex systems}
\author{G Benk{\"o} and H J Jensen}
\address{Department of Mathematics, Imperial College London,\\
South Kensington Campus, SW7 2AZ, London, UK}
\eads{\mailto{g.benkoe@imperial.ac.uk}, \mailto{h.jensen@imperial.ac.uk}}

\begin{abstract}
For large coupled nonlinear systems, it is difficult to visualize the high-dimensional phase space, which has been thoroughly studied in smaller systems with regards to phenomena such as riddled basins. Here we propose a method to reduce the phase space by defining a phase space cross section. The method is applied to a system of dynamically coupled maps introduced by Ito~\&~Kaneko (\textit{Phys.\ Rev.\ Lett.}, 88, 028701, 2001 \& \textit{Phys.\ Rev.\ E}, 67, 046226, 2003).
We show that the transitions between phases of different synchronization behaviour are not always sharp but can be characterized by fractal boundaries in both phase and parameter space. 
\end{abstract}


\section{Introduction}
A hallmark of chaos in complex systems is sensitive dependence on initial conditions. An important property of the phase space of complex systems are basins of attraction, describing the sensitivity of the final state of a system depending on the initial conditions. In the case of riddled basins, multiple basins are present, intermingling in a fractal way and exarcebating this sensitivity \cite{Alexander1992,Ashwin1996}. However, properties of the phase space are difficult to study for large coupled nonlinear systems, as the phase space is  high-dimensional and hard to visualize.

For instance, in this paper we are interested in studying large-scale synchronization, a dynamical property of networks that is widely observed in nature, for instance in the brain \cite{Gray1989,Varela2001}. Synchronization has been analyzed for many physical systems \cite{Pecora1997,Pikovsky2001}, one model for synchronization are globally coupled maps (GCM)\cite{Ito2001,Kaneko2000}, where the individual maps constituting GCM have been extensively studied in dynamical systems research, especially the logistic map.

Here we propose a method to reduce the phase space of such large systems by defining a phase space cross section, allowing us to apply methods for the analysis of dynamical systems to large GCM. The aim is also to gain a better understanding of large network dynamics such as synchronization through these methods.

\section{Model}

We consider a GCM introduced by Ito~\&~Kaneko \cite{Ito2001,Ito2003}, where not only the maps but also the couplings between them are dynamical variables. This is closer to real-world systems, where connections are not always static but have their own dynamics. Thus in the GCM, there are a set of variables ${x^i}$, called units in the following, forming a network with connections of variable weights $w^{ij}$, related by

\begin{eqnarray}
x^i_{n+1} & = & \left(1-c\right) f(x^i_n) + c \sum^N_{j=1} w^{ij}_n f(x^j_n) \label{eq:gcm1} \\
f(x) & = & a x \left(1-x\right) \\
w^{ij}_{n+1} & = & \frac{[ 1 + \delta g(x^i_n,x^j_n) ] w^{ij}_n}
		{\sum^N_{j=1} [ 1 + \delta g(x^i_n,x^j_n) ] w^{ij}_n} \\ 
g(x,y) & = & 1 - 2 \left|x-y\right| \,.
\label{eq:gcm4}
\end{eqnarray}

$a$ is the logistic equation parameter,
and $c$ is the coupling parameter.
The function $g$ defines a Hebbian update of the connection weights, by reinforcing the
connections between similar units. It is scaled by $\delta$.

\begin{figure}\centering
\includegraphics[width=.7\textwidth]{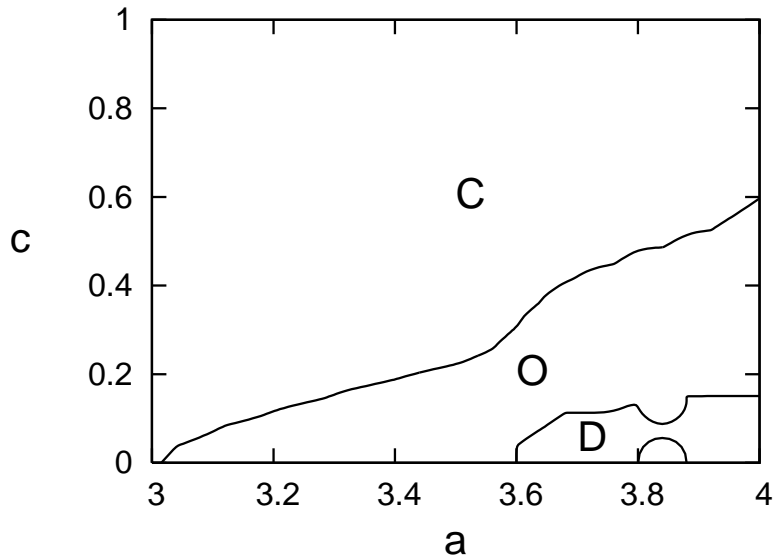}
\caption{Schematic phase diagram of the GCM with $N = 100$ depending on parameters $a$ and $c$, averaged over 500 samples. There are three phases: the coherent phase $C$, ordered phase $O$, and disordered phase $D$. After \protect\cite{Ito2001}.}
\label{fig:pd}
\end{figure}

The system exhibits three different long-term behaviours which can be classified into three phases, depending on the parameters. 
In the coherent phase, all the units synchronize, forming one synchronized cluster containing all the units.
In the ordered phase, the set of units is partitioned into subsets or clusters $C_k$ within which there is synchronization or which contain single units not synchronized with any other unit. In line with \cite{Ito2001,Ito2003}, we will call the number of parts in this partition in the following the number of synchronized clusters. Finally, in the disordered phase, no synchronization at all is achieved, forming a partition with $N$ synchronized clusters. The phase diagram in figure~\ref{fig:pd} shows the boundaries between predominant phases in the parameter space. We are especially interested in the phase transitions across the boundaries \cite{Peel2007} and will study them further in the next sections.

The connection strength $w^{ij}$ between units in synchronized clusters $C_k$ is around $1/N_{C_k}$. The connection strength between units in different synchronization clusters is vanishing \cite{Ito2003}. This fact allows us to easily identify synchronized clusters.

\section{Phase space cross section}

We next examine this system using nonlinear systems methods. In \cite{Pecora1997}, Pecora {\em et al.}\ present some studies on the synchronization of small chaotic systems based on stability analysis and bifurcation theory. 
For instance, an intriguing behaviour observed in GCM are riddled synchronization attractor basins. A basin is riddled when for every point in the basin a small error might lead to a different attractor, the two attractors are completely intermingled \cite{Alexander1992,Ashwin1996}. This is related to the concept of fractal basin boundaries \cite{McDonald1985,Takesue1984}, for which it suffices that the borders of two attractor basins are intermingled. This phenomenon is especially relevant and important regarding the final state of the system. The final state in physical systems with riddled basins is uncertain around these basin boundaries \cite{Ott1994,Pecora1997}.

However, we are interested in large systems, where it is not practicable to visualize the phase space in order to gain insight into its characteristics. Typically, in our studies the GCM is composed of 100 units. 

\begin{figure}
\centering
\subfigure[]{\includegraphics[width=.4\textwidth]{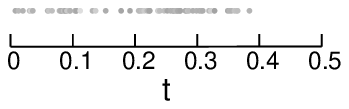}\label{fig:p}}
\subfigure{\includegraphics[width=.1\textwidth]{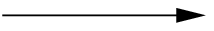}}
\setcounter{subfigure}{1}
\subfigure[]{\includegraphics[width=.35\textwidth]{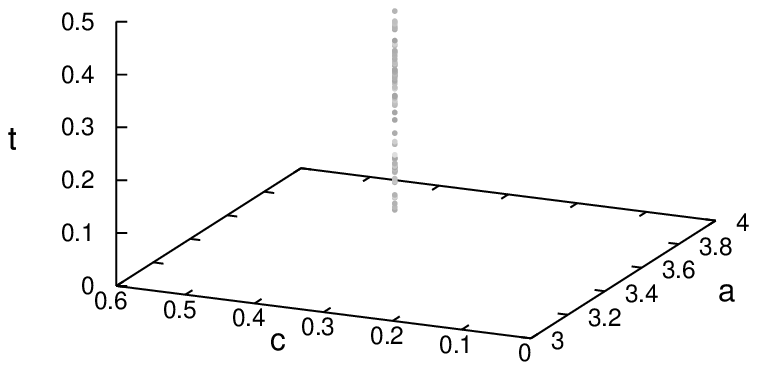}\label{fig:p3d}}\\
\subfigure[]{\includegraphics[width=.8\textwidth]{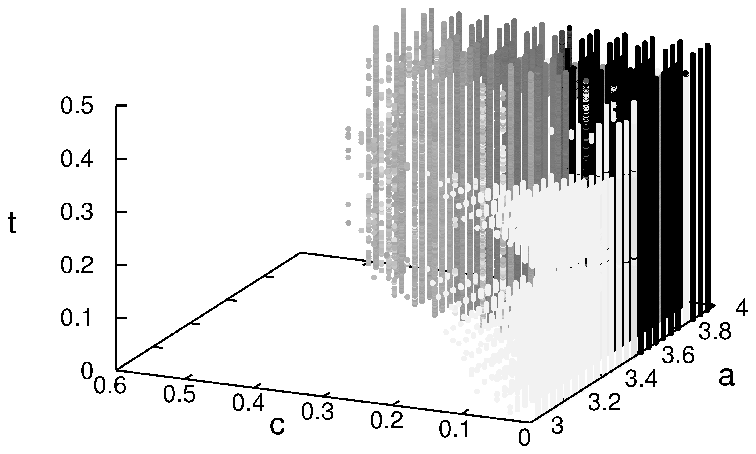}
\includegraphics[width=.1\textwidth]{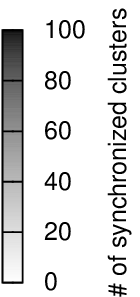}\label{fig:pd3d}}
\caption{Expanding the phase diagram by the phase space cross section. (a) The phase space cross section for $a = 3.76$ and $c = 0.40$. 
The number of synchronized clusters is plotted against the parameter $t$ of (\ref{eq:curv}), going from white (coherent) over gray shades (ordered) to black (disordered phase). 
(b) The same phase space cross section, 
placed along the $t$ axis in the space spanned by $(a,c,t)$ on the line defined by $a = 3.76$ and $c = 0.40$. 
(c) Repeating the plotting of the previous phase space cross section for and at every $(a,c)$, thus obtaining a phase diagram in the space spanned by $(a,c,t)$.}
\end{figure}

We thus aim to reduce the phase space to make it amenable to our studies. 
Our approach is to consider a one-dimensional curve in phase space as representative of the phase space. The shape of the curve should take advantage of the symmetries of the system defined by (\ref{eq:gcm1}) to (\ref{eq:gcm4}) and avoid the borders and diagonals of the phase space, where there is trivial synchronization.
We used the parametric curve $P$ defined by
\begin{multline}
P = \left\{ \left(x_0^0 = \frac{1-\cos{\pi t}}{2},\dots
	     x_0^i = \frac{i}{N-1} \frac{\sin{\pi t}}{2} +
	     \frac{N-1-i}{N-1} \frac{1-\cos{\pi t}}{2},\dots   \right. \right. \\
	    \left. \left. x_0^{N-1} = \frac{\sin{\pi t}}{2} \right) \in \left. \mathbb{R}^N \right| t \in [0,0.5]\right\}\,, 
\label{eq:curv}
\end{multline}
calculating the outcome of simulating (\ref{eq:gcm1}) to (\ref{eq:gcm4}) as we change the initial conditions along $P$ and keeping all control parameters ($a$, $c$, $\delta$) constant.
We characterize the outcome by the number of synchronized clusters.
The variation of the number of synchronized clusters along $P$ thus yields a one-dimensional cross section of the phase space, i.e.\ a summary of the appearance of the corresponding phase space.

An example of how we will represent the behaviour along $P$ is shown for $a = 3.76$ and $c = 0.40$ in figure~\ref{fig:p}. The number of synchronized clusters is plotted with dots of different gray values against $t \in [0,0.5]$, increasing with the number of synchronized clusters, from white (coherent) over gray (ordered) to black (disordered phase). 

\section{Results and discussion}

\begin{figure}\centering
\includegraphics[width=.5\textwidth]{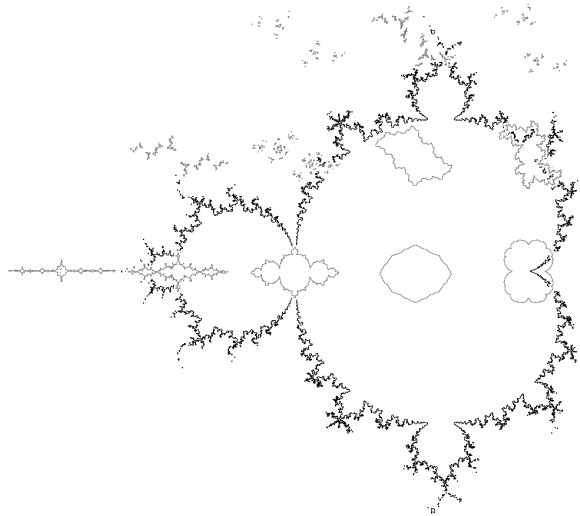}
\caption{An ``extended phase diagram'' of the complex map $f_c(z) = z^2+c$. By superimposing on a sample of $c$ values of the Mandelbrot set the corresponding Julia set, an analogous diagram to figure~\ref{fig:pd3d} is obtained.}
\label{fig:mj}
\end{figure}

Using the previously defined phase space cross section, we can now study bigger systems, and extend the study of fractal basin boundaries to large GCM. For example, we can use the phase space cross section $P$ of figure~\ref{fig:p} and calculate the fractal dimension of the boundary between the coherent and ordered regions in $P$, thus obtaining an insight into the nature of the boundary in the complete phase space. For example, the boundary within $P$ shown in figure~\ref{fig:p} is Cantor set-like and has a box-counting dimension of $D_0 \approx 0.45$\,.

We will use our phase space cross section in order to obtain a more detailed phase diagram of the GCM. 
We expand the GCM's phase diagram shown in figure~\ref{fig:pd}, which is drawn in parameter space, by an additional dimension constructed from phase space, the phase space cross section. Figure~\ref{fig:p3d} shows a phase space cross section plotted perpendicularly to the $(a,c)$ plane over the $a$ and $c$ parameter values it has been calculated for. In figure~\ref{fig:pd3d}, we repeat this for each pair $(a,c)$ and add all the phase space cross sections perpendicularly to the $(a,c)$ plane over their corresponding $(a,c)$ values 
to obtain a more detailed phase diagram.

This process can be nicely illustrated by imagining it for the complex map $f_c(z) = z^2+c$. In this case the phase space is just two-dimensional, spanned by $\textrm{Re}(z)$, $\textrm{Im}(z)$, and the behaviour in phase space is described by a Julia set, varying for each $c$. The parameter space is also two-dimensional, spanned by $\textrm{Re}(c)$, $\textrm{Im}(c)$, and the behaviour is now described by the Mandelbrot set. The analogy to the more detailed phase diagram is obtained by adding a sketch of the corresponding Julia set to each or a sample of $c$ values on top of a Mandelbrot set, see figure~\ref{fig:mj}.

\begin{figure}\centering
\includegraphics[width=.8\textwidth]{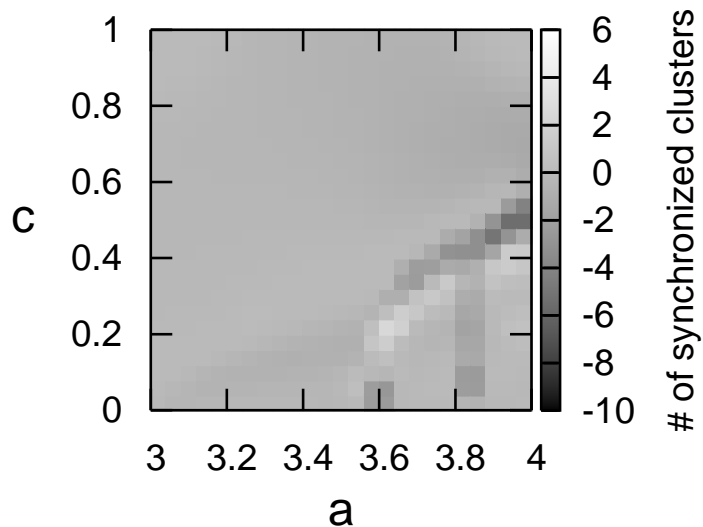}
\caption{Difference in number of synchronization clusters between phase diagrams obtained by random sampling and using $P$.}
\label{fig:pddiff}
\end{figure}

The two-dimensional and three-dimensional phase diagrams are similar, most of the parameter space is occupied by the coherent phase, while there are transitions into a first ordered, then disordered phase as $a$ increases and $c$ decreases. To appreciate how representative of the whole phase space $P$ is, we can average over the number of synchronized clusters in $P$ for each $(a,c)$ and compare it to the average over a random sample for each $(a,c)$. The result is shown in figure~\ref{fig:pddiff}. The difference in the phase diagrams obtained by random sampling and using $P$ is indeed small compared to $N=100$.

We can now see that the ordered phase is split into a part I below $a \approx 3.57$, which is the onset of chaos in the logistic map, and a part II above it.
In the former part, the behaviour of the $N = 100$ system is characterized by 2--6 synchronized clusters while in the latter part the system typically decomposes into 30--50 synchronized clusters. Also, the added dimension of the new phase diagram reveals that the boundaries of the phases are not simply smooth. 
The transition $C/O_I$ is sharp, there is not an intermittent but a sudden change of phase at a threshold $(a,c)$, which nevertheless varies smoothly along the phase space cross section.

\begin{figure}\centering
\includegraphics[width=.8\textwidth]{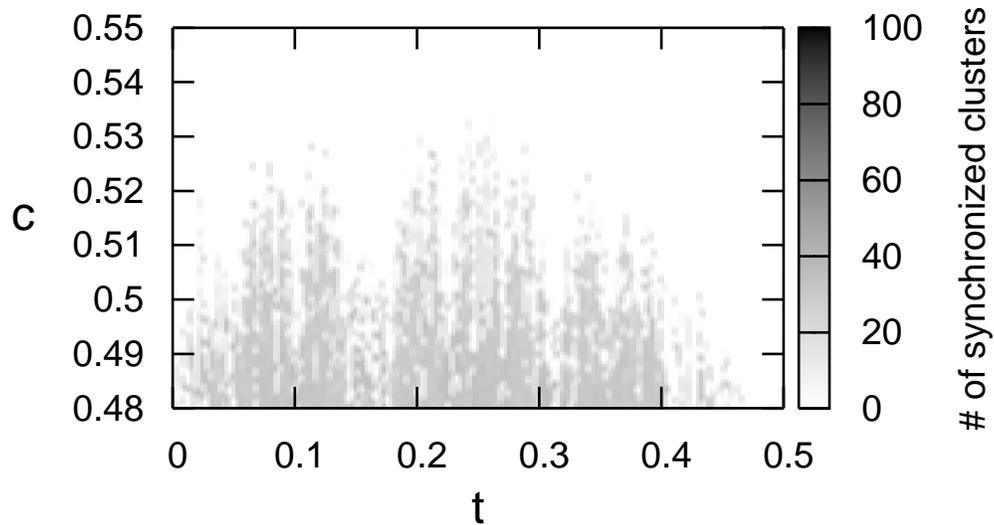}
\caption{Plot of the phase space cross section as a function of $c$.}
\label{fig:xc}
\end{figure}

However, the boundary $C/O_{II}$ seems to be fractal in both phase and parameter space. For more clarity, we can draw the phase space cross section against the parameter $c$, with $a=3.97$ for example, see figure~\ref{fig:xc}. We observe indeed that as $c$ increases from 0.45 to 0.55, the phase space cross section changes from being completely in the ordered to completely in the coherent phase, going through a range where the cross section is alternating between the two phases in a fractal way, indicating a fractal basin boundary \cite{McDonald1985,Takesue1984} in phase space.
This is corroborated by finding more fine structure for higher plot resolutions. 

\begin{figure}\centering
\subfigure[]{\includegraphics[width=.45\textwidth]{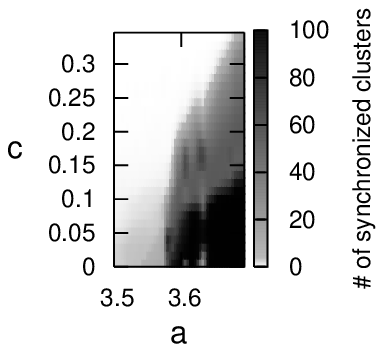}\label{fig:fpavg}}
\subfigure[]{\includegraphics[width=.45\textwidth]{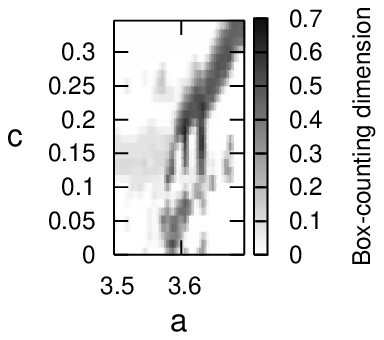}\label{fig:fp}}
\caption{(a) Detail of the phase diagram of the GCM. (b) Box-counting dimension of the boundaries between regions $C$, $O$, and $D$ within the phase space cross section for each $(a,c)$ in the phase diagram (a).}

\end{figure}

As explained above, we can quantify the fractal nature of the boundaries in phase space 
by calculating the box-counting dimension $D_0$ of the boundary within the phase space cross section \cite{McDonald1985}. Figure~\ref{fig:fpavg} shows a detail of the phase diagram where the regions $C$, $O_{I}$, $O_{II}$, and $D$ meet, around $a = 3.6$ and $c = 0.1$. Next to it, figure~\ref{fig:fp} shows the fractal dimension $D_0$ of the boundary between $C$, $O_{I}$, $O_{II}$, and $D$ within the phase space cross section for each pair $(a,c)$ in the considered part of the phase diagram. By comparing the phase and fractal dimension diagrams, we see that within the regions $C$, $O_{I}$, $O_{II}$ of the phase diagram, $D_0 = 0$. Indeed, there the phase space and thus the phase space cross section are ``filled'' with one single region and contain no boundary. Where $C$ and $O_I$ meet, $D_0$ is also near 0. There the boundary within phase space cross section consists of a few points, and thus has theoretical dimension of 0. However, where the regions $C$ and $O_{II}$ meet, $D_0 \approx 0.5$\,, quantifying the fractal nature of this border. The change of the $C/O$ border from sharp ($C/O_I$) to fractal ($C/O_{II}$) is called a boundary metamorphosis \cite{Grebogi1987a}.

Most of the border $O/D$ has $D_0 \approx 0$ and is not fractal, but for $a < 3.67$ $D_0$ increases up to values of 0.5\,. There is a lot of detail in this region which will be studied in future work. 

The fractal nature of the boundaries in parameter space can be quantified as well, by calculating the box-counting dimension of the boundaries within the phase diagram and averaging over a sample of initial conditions. Thus if we calculate the dimension looking only at the boundaries $C/O_I$, $C/O_{II}$, $O_I/D$, and $O_{II}/D$, we obtain almost always values near 1, except for $C/O_{II}$ where $D_0 \approx 1.6$. Thus $C/O_{II}$ is fractal in parameter space. As it was also the only fractal boundary in phase space, this indicates a correlation between the fractal nature in phase and parameter space, which has been partly proven for the complex map $f_c(z) = z^2+c$ \cite{Lei1990}.

\begin{figure}\centering
\includegraphics[width=.7\textwidth]{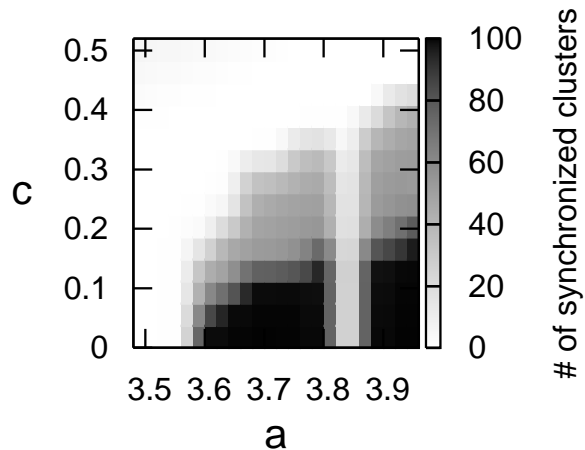}
\caption{The slice corresponding to $t = 0.499$ of the extended phase diagram in figure~\ref{fig:pd3d}.}
\label{fig:bord}
\end{figure}

Finally, a feature of the extended phase diagram in figure~\ref{fig:pd3d} is the small mixed coherent/ordered enclaves in the otherwise disordered phase, for example for $t \approx 0.5$, and $a \approx 3.85$, also seen in the slice, shown in figure~\ref{fig:bord}, of the extended phase diagram corresponding to $t = 0.499$. 
This seems to indicate the advantage of using the phase space cross section defined by (\ref{eq:curv}), as follows. For $t \approx 0$ or $t \approx 0.5$, i.e.\ the beginning and end of $P$, the $x_0^i$ values are close to each other, with $x_0^i \approx 0$ or $x_0^i \approx 0.5$ respectively. This allows probing into the behaviour of the system for almost synchronized initial conditions. Thus the small mixed coherent/ordered enclaves in the otherwise disordered phase correspond to the islands of periodicity within the chaotic regime of an individual logistic map at $a \approx 3.85$, $a \approx 3.7$,\dots Thus we have an example where for almost synchronized initial conditions, the behaviour of the GCM is strongly influenced by the characteristics of the individual maps $x^i$. 

\section{Conclusion}

In conclusion, we have constructed a tool for efficiently probing and further understanding the dynamics of networks of coupled maps, in this case the transitions between states of synchronization in a GCM.
The phenomenon of fractal boundaries in phase and/or parameter space was found using this tool in the studied model, as suggested by previous findings in similar models \cite{Lai1994}, reviewed in \cite{Pecora1997}.
One example for the relevance of the synchronization observed in GCM are neural networks.
There is evidence that the elementary cognitive acts underlying cognition are achieved by transient neural assemblies dynamically linked by synchronization \cite{Varela2001}. Studying phases is thus important in the context of analyzing mental states.
For example, a riddled phase space might potentially facilitate switching between disordered dynamics in a neural network and the emergence of synchronized assemblies.

\ack{}
Useful discussions with Adele Peel, computer support by Andy Thomas and the use of the Imperial College High Performance Computing Service 
are gratefully acknowledged.

\section*{References}

\bibliographystyle{jphysicsB}
\bibliography{dynsync}

\end{document}